\documentclass[10 pt,twocolumn]{article}       
\usepackage{latexsym}
\usepackage{graphicx}

\usepackage{amsthm}
\usepackage{amssymb}   
\usepackage{epsfig}
\usepackage{amsmath}
\usepackage{color}
\usepackage{authblk}

\date{}
\begin{document}

\title{Quasi-2D $J_1$-$J_2$ Antiferromagnet $Zn_2VO(PO_4)_2$ and its $Ti$-substituted Derivative - A Spin-wave Analysis} 
\author{Satyaki   Kar{\footnote{CMP Div., Saha Institute of Nuclear Physics, 1/AF Salt Lake, Kolkata-64, India.}}   and   Tanusri Saha-Dasgupta{\footnote{S. N. Bose National Centre for Basic Sciences, Block-JD, Salt Lake, Kolkata-98, India.}}  }  

\maketitle

\begin{abstract}
In this study, we present non-linear spin wave analysis of a quasi 2D spin-$\frac{1}{2}$ $J_1-J_2$ antiferromagnet at the parameter regime relevant for the recently studied compound $Zn_2VO(PO_4)_2$. We obtain the temperature dependence of the spin wave energy, susceptibility and magnetization using Green's function technique and Hartree-Fock factorization or Tyablikov's decoupling approximation. The comparison of our numerical results with the experimental findings is discussed. Magnetic structure factor is calculated and compared with powder neutron diffraction data. We also study the spin wave behavior of the compound $Zn_2Ti_{0.25}V_{0.75}O(PO_4)_2$ obtained by  partial chemical substitution of $Ti$ at $V$ sites of the compound $Zn_2VO(PO_4)_2$ (Phys. Rev. B87, 054431). Due to the superlattice structure of the spin lattice, the substituted compound possess multiple spin wave modes. The spin wave analysis confirms the quasi-1D nature of the substituted system.

\end{abstract}

\section{Introduction}
Over the past several years, low dimensional spin systems, belonging to family of cuprates\cite{ref1}, vanadates\cite{ref2}, titanates\cite{ref3} 
have emerged as a field of active research in the condensed matter community. 

In this context, the magnetic properties of the layered vanadium compounds $AA^\prime VO(PO_4)_2$ ($A, A^\prime=Pb,Zn,Sr,Ba$) has drawn interest as they are found to be quasi two dimensional (2D) in nature with in-plane spin-frustration that corresponds well to a $J_1-J_2$ Heisenberg antiferromagnetic (AF) model\cite{thalmeier}. Early transition metal vanadium ($[Ca]3d^34s^2$) is in its nominal 4+ valences in the compounds giving rise to the possibility of realizing the spin 1/2 2D AF. 
 The compound $Zn_2VO(PO_4)_2$, in particular, has attracted attention due to interesting and unusual magnetic properties\cite{Lii,kini,yusuf}, showing weak superexchange ($\sim$ meV) mediated low dimensional magnetic behavior of V spins. Susceptibility and specific heat measurements\cite{kini} and neutron diffraction results\cite{yusuf} have confirmed the quasi 2D spin-$\frac{1}{2}$ AF nature for the spin lattice of the system. In spite of existence of a 3D Neel temperature of $T_N=3.75$ K, the compound shows signatures of short range 2D magnetic correlations well up to 6.95 K\cite{yusuf}. The broad peaks observed in magnetic-diffraction pattern of the compound, arising from the diffuse scattering at low temperatures ($T<7~K$) have been attributed partly to spin waves (for $T<T_N$) and partly (for $T>T_N$) to short range 2D magnetic correlations.
 In this paper, we represent a spin-wave analysis of the problem at low temperature taking into account the spin fluctuations within the Tyablikov's decoupling approximation (TDA)\cite{tyablikov,lines,pat,haar} on top of the mean-field linear spin wave (LSW) results and study the temperature dependence of the magnetic properties such as magnon dispersion, magnetization and spin susceptibility. Our calculation at mean-field+TDA level shows the spin waves to exist up to a temperature of $T_{max}=4.31$ K.

A recent density functional theory (DFT) based study\cite{sudipta} reported partial substitution of vanadium ions by titanium ions in $Zn_2VO(PO_4)_2$, resulting in interesting modification of the crystal structure especially for the Ti concentration of $x=0.25$. The underlying spin lattice for such compound, $Zn_2Ti_{0.25}V_{0.75}O(PO_4)_2$, turns out to be quasi-1D AF with weak couplings in both the transverse directions. In our present study we also carry out spin wave analysis for this Ti-substituted compound and present a comparative study for the spin-wave modes obtained for the pristine and substituted compounds. Our calculation on the substituted compound produces three different magnon modes.

The content of our paper is arranged as follows. In Section 2 we discuss the general formulation for the non-linear spin wave analysis of a quasi-2D system with frustration. In section 3, we report our numerical results for $Zn_2VO(PO_4)_2$ and compare with the experimental findings. Section 4 is devoted to the compound $Zn_2Ti_{.25}V_{.75}O(PO_4)_2$ which includes a linear and TDA-level spin wave study of this quasi 1D spin 1/2 system. Section 5 discusses the comparison of magnon spectra of the two systems and Section 6 gives the summary of our work.

\section{Formulation}
\begin{figure}[htp]
\begin{center}
\epsfig{file=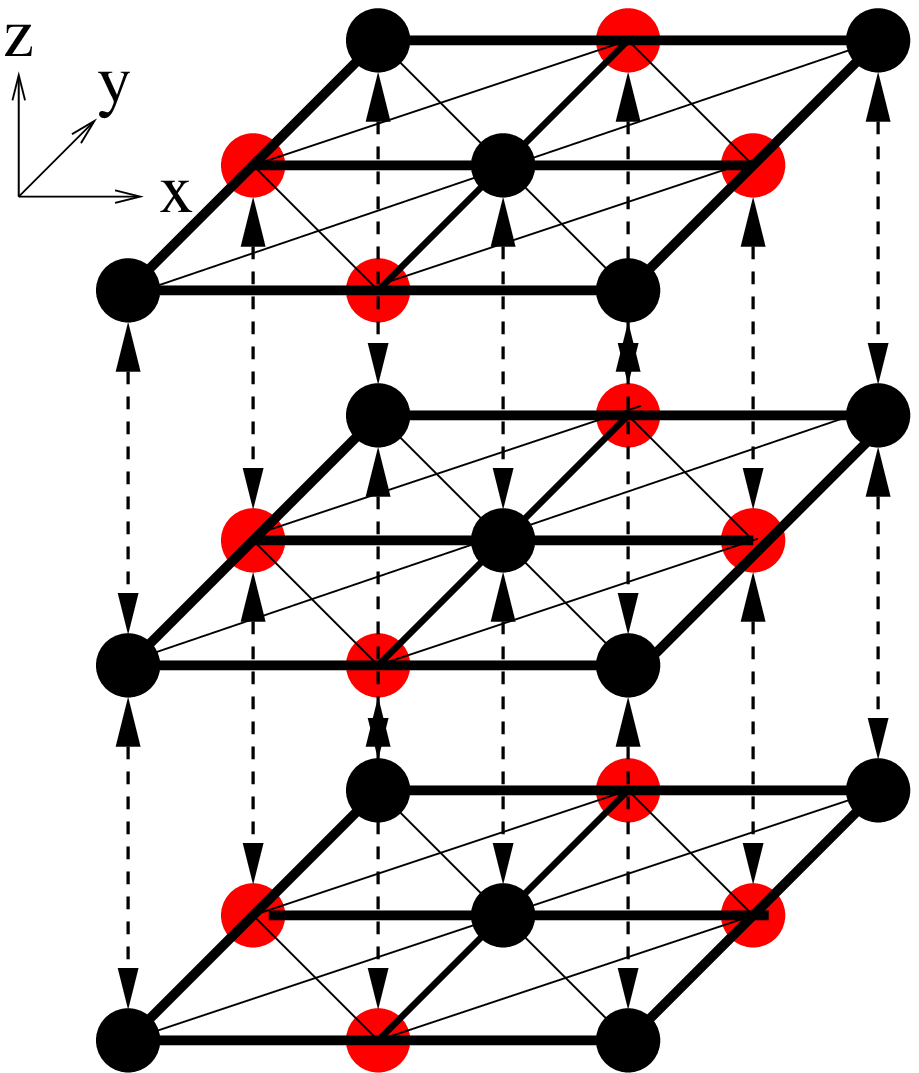,width=.47\linewidth,clip=}\hskip .3 in
\epsfig{file=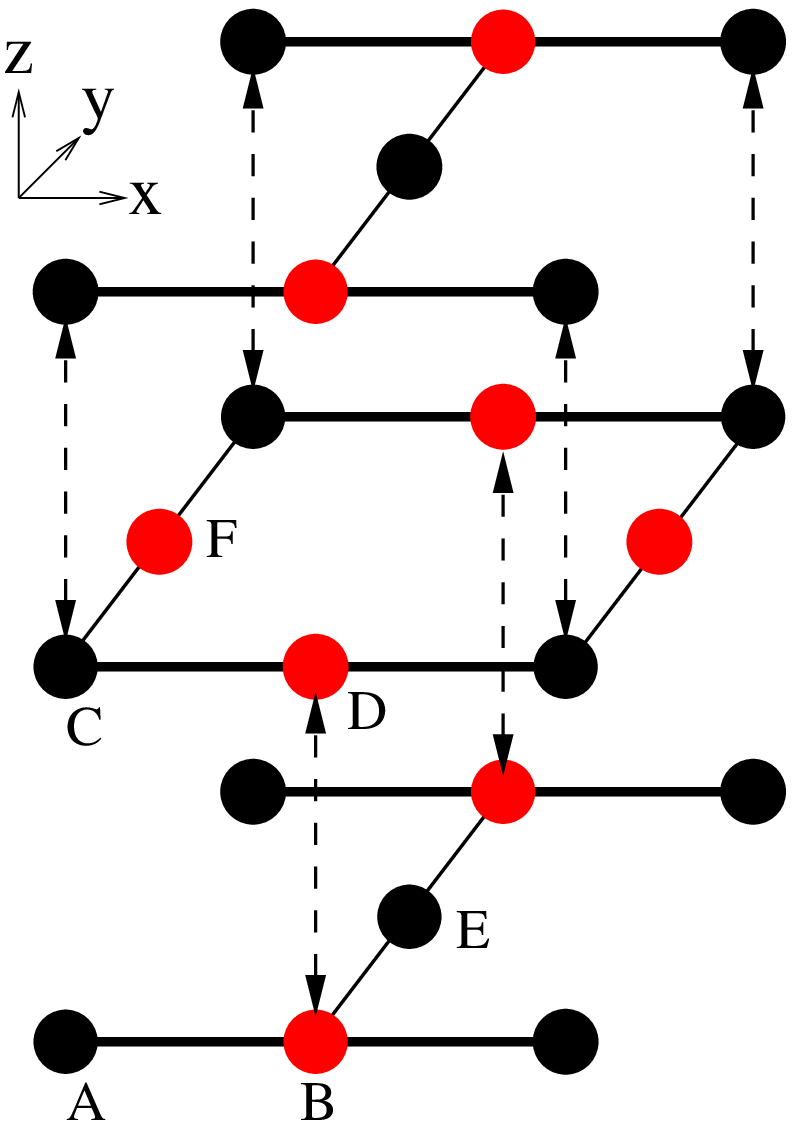,angle=0,width=.4\linewidth,clip=}
\end{center}
\caption{(Color online) The AF spin reference state for $Zn_2VO(PO_4)_2$ (left) and  $Zn_2Ti_{.25}V_{.75}O(PO_4)_2$ (right). Crystal sites are colored black and red (gray) alternately to represent the up and down spins respectively. Left: The thick solid, thin solid and dashed lines represent in-plane $J_1$ and $J_2$ and out-of-plane $J_c$ interactions respectively. Right: The thick solid, thin solid and dashed lines represent strongest $J_1^\prime$ and weak $J_1^{\prime\prime}$ and $J_c^\prime$ interactions respectively. A, B, C, D, E, F are the six sublattices of the spin lattice (see Sec IV for details).}
\label{fig1}
\end{figure}

$Zn_2VO(PO_4)_2$ has a tetragonal crystal structure of space group I4cm with lattice parameters a = 8.93 $\AA$ and c = 9.04 $\AA$\cite{sudipta}.
The spin model for the compound $Zn_2VO(PO_4)_2$ is a quasi-2D spin-1/2 Heisenberg quantum antiferromagnet in a cubic lattice with magnetic $V^{4+}$ ions sitting at the lattice sites. The $xy$ plane contains the prominent nearest neighbor (NN) antiferromagnetic coupling along with relatively weaker next nearest-neighbor (NNN) AF coupling which are stacked along the $z$-direction coupled via weak ferromagnetic (FM) coupling\cite{yusuf} (Fig.\ref{fig1}, left panel). Starting from this AF reference state, we carry out a spin wave study for the system and examine the magnetization, susceptibility and the extent of Neel ordering. 
Holstein-Primakoff transformation is used to obtain the magnon modes in LSW approximation. In the next step, we extend our calculation beyond this approximation accounting for the additional terms of the Hamiltonian that are quadratic in spin-wave operators. In order to do that we use decoupling approximation\cite{haar} of the non-linear terms and solve the equation of motion of the magnon Green's function yielding the improved expression for the spin-wave modes.

The overall Hamiltonian is, thus, given by
\begin{align}
H&=J_1\sum_{i,\delta}{\bf{S_i}.{S_{i+\delta}}}+J_2\sum_{i,\delta^\prime}{\bf{S_i}.{S_{i+\delta^\prime}}}-J_{1c}\sum_{i,\delta_c}{\bf{S_i}.{S_{i+\delta_c}}}.
\label{eq1}
\end{align}
with $J_{1c}=\lambda J_1$ and $J_2=\alpha J_1$. $\delta$ ($\hat{x},\hat{y}$) and $\delta^\prime$ ($\hat{x}+\hat{y},\hat{x}-\hat{y}$) denote NN and NNN lattice positions in $xy$ plane respectively whereas $\delta_c$ gives NN lattice positions along $c$ or $z$ directions.

Rewriting the Hamiltonian in terms of spin wave operators (about a reference state which is AF in $xy$ plane and FM along $z$ direction) and retaining terms up to quadratic in bosonic spin wave operators the Fourier transformed Hamiltonian is given by
\begin{eqnarray}
{H}&=&E_{0}+4J_1S\sum_{k}[g_{11}(k)(a_k^\dagger a_k+b_k^\dagger b_k)+\nonumber\\&&g_{12}(k)(a_kb_{k}+h.c.)]
\label{eq2}
\end{eqnarray}
where $a_k,~b_k$ are Fourier transformed bosonic spin-wave operators\cite{stratos,pratap}, $g_{11}(k)=1+\alpha(\Gamma_k-1)+\lambda(1-cos~k_z)/2$, $g_{12}(k)=\gamma_k$=(cos $k_x$+cos $k_y$)/2, $\Gamma_k$=cos($k_x$)cos($k_y$) and $E_{0}$ is the vacuum or ground state energy. The sum is over the first Brillouin Zone (BZ).

Next we consider the non-linear correction to our results where additional terms that are quadratic in $S^z$ or quartic in bosonic spin wave operators are considered as well. The overall correction becomes
\begin{align}
&{H}_2=\frac{-8J_1}{N}a_k^\dagger a_{k^\prime}[\gamma_{k-k^\prime}b_{k+q}^\dagger b_{k^\prime +q}+\frac{\alpha~cos(k_{z}-k^\prime_{z})}{2}\nonumber\\&a_{k^\prime+q}^\dagger a_{k+q}]+\frac{4J_2}{N}\Gamma_{k-k^\prime}[a_{k}^\dagger a_{k^\prime}a_{q-k}^\dagger a_{q-k^\prime}+b_{k}^\dagger b_{k^\prime}b_{q-k}^\dagger b_{q-k^\prime}]\nonumber\\&-\frac{2J_2}{N}\Gamma_{q+k^\prime}[a_k^\dagger a_{k^\prime}a_{k+q}a_{k^\prime+q}^\dagger +b_k^\dagger b_{k^\prime} b_{k+q}b_{k^\prime+q}^\dagger+h.c.]\nonumber\\&\frac{-2J_1}{N}[a_{k^\prime -q}^\dagger a_{k^\prime}a_{k}(\gamma_{k+q} b_{k+q}+\frac{\alpha~cos(k_{z}+q_{z})}{2}a_{k+q}^\dagger)\nonumber\\&+(\gamma_{k}a_{k}+\frac{\alpha~cos(k_{z})}{2}b_{k}^\dagger)b_{k^\prime}^\dagger b_{k+q}b_{k^\prime -q} + h.c.]
\label{eq3}
\end{align}
where sum over $k,k^\prime$ and $q$ are implied.

In order to find the spin wave dispersions and susceptibility, we first obtain the equation of motion\cite{lines,haar,callen} of the bosonic Green's functions, 
\begin{align}
G_{11}(k,\omega)&=<<a_k|a_k^\dagger >> , G_{22}(k,\omega)=<<b_k|b_k^\dagger >>\nonumber\\G_{12}(k,\omega)&=<<b_k^\dagger|a_k^\dagger >> , G_{21}(k,\omega)=<<a_k|b_k >>\nonumber
\end{align}
where $<<A|B>>$ denotes the Fourier transform of the Green's function involving operators A and B.
Fluctuation (nonlinear) terms are averaged using TDA and the resulting equations of motion become
\begin{align}
\omega G_{11}(k,\omega)&=\frac{1}{2\pi}+a_{11}G_{11}(k,\omega)+a_{12}G_{12}(k,\omega)\nonumber\\
-\omega G_{12}(k,\omega)&=a_{21}G_{11}(k,\omega)+a_{22}G_{12}(k,\omega)
\label{eq4}
\end{align}
where $a_{11}=a_{22}=4J_1S[{g}_{11}(k)+\zeta_1(k)]$, $a_{12}=a_{21}=4J_1S[g_{12}(k)+\zeta_2(k)]$, $\zeta_1(k)=\zeta_2(0)-\frac{1}{NS}[-2\alpha\Gamma_k+\sum_q(\Gamma_k+\Gamma_{q+k})<\eta_q>]$ and $\zeta_2(k)=-\frac{2}{NS}\sum_q[g_{12}(k)<\eta_q>+g_{12}(k-q)<\xi_q>]$  and $<\eta_q>=<a_q^\dagger a_q>=<b_q^\dagger b_q>$, $<\xi_q>=<a_qb_q>=<b_q^\dagger a_q^\dagger>$ are the expectation values \cite{pratap}.
From the solution of Eq.\ref{eq4} we get\cite{lines}
\begin{equation}
4\pi G_{11}(k)=\frac{1-A}{\omega+\omega_k}+\frac{1+A}{\omega-\omega_k}~~~~~~~~~
\label{eq5}
\end{equation}
where $\omega_k$, being the pole of the Green's function, is the spin wave dispersion which is given by
\begin{equation}
\omega_k=4J_1S[{g}_{11}(k)+\zeta_1(k)]\sqrt{1-f(k)^2}
\label{eq6}
\end{equation}
$f(k)=\frac{g_{12}(k)+\zeta_2(k)}{g_{11}(k)+\zeta_1(k)}$ and $A=\frac{1}{\sqrt{1-f(k)^2}}$. The spectral representation of correlation functions then leads to
\begin{eqnarray}
&&<\eta_k>=1/2[A~coth(\beta\omega_k/2)-1]\nonumber\\
&&<\xi_k>=-1/2f(k)A~coth(\beta\omega_k/2)
\label{eq7}
\end{eqnarray}
which are actually thermal averages in a canonical ensemble (expectation value).
\begin{figure}[h]
\begin{center}
\epsfig{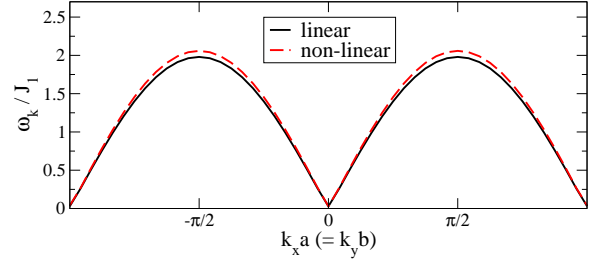}
\end{center}
\caption{Magnon dispersion of $Zn_2VO(PO_4)_2$ along $k_x=k_y$ direction (and $k_z=0$) at zero temperature with linear and non-linear approximation. ``a'' and ``b'' are lattice constants along $x$ and $y$ directions.}
\label{fig2}
\end{figure}
\begin{figure}[h]
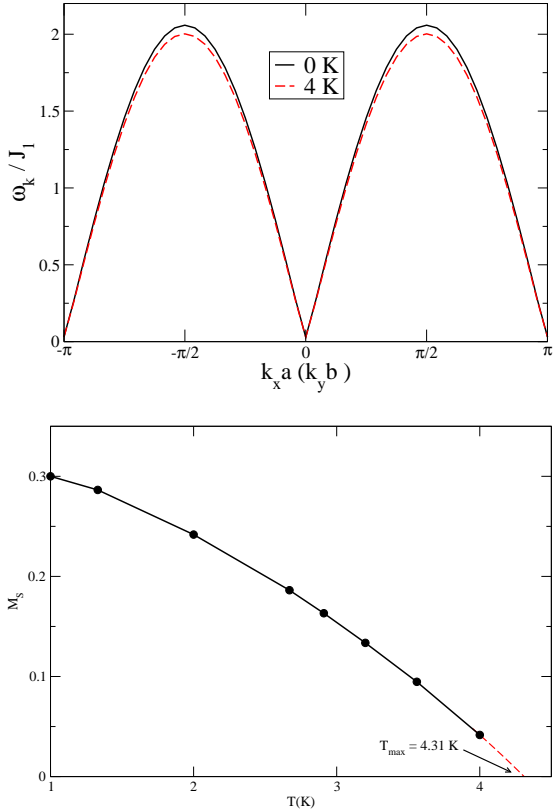

\begin{center}
\epsfig{file=nonl-spwv-disp3_new.eps,width=.9\linewidth,clip=}\\\vskip .15 in
\epsfig{file=mag2_new.eps,width=.9\linewidth,clip=}
\end{center}
\caption{(Top) Magnon modes along $k_x=k_y,k_z=0$ at various temperatures using non-linear approximation. (Bottom) sublattice magnetization vs. temperature plot for the quasi-2D AF structure.}
\label{fig3}
\end{figure}

The sublattice magnetization is given by 
\begin{equation}
M(T)=S-N^{-1}\sum_k[A~coth(\beta\omega_k/2)-1].
\label{eq8}
\end{equation}
The k-vector summation runs over the AF BZ only.

\begin{figure}
\begin{center}
\epsfig{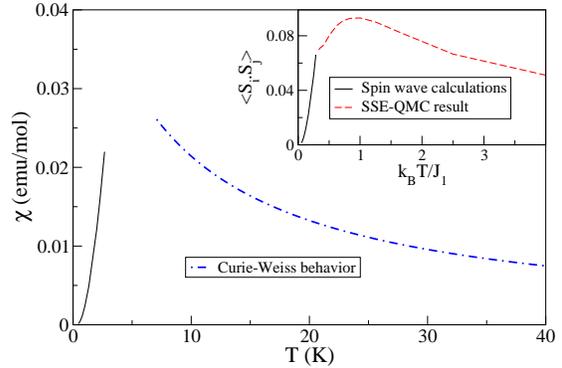}
\end{center}
\caption{Parallel susceptibility vs. temperature along with the high temperature Curie-Weiss behavior. The inset shows the spin-correlations using spin-wave results and SSE-QMC results.}
\label{fig4}
\end{figure}
\section{Numerical Results}
Calculations were carried out on a $48\times48\times4$ lattice, which is justified by the quasi 2D nature of the system. We used the parameter values $\alpha=0.02$ and $\lambda=0.03$ to represent the compound $Zn_2VO(PO_4)_2$. The strength of the strongest coupling used is $J_1=8~K$\cite{sudipta}. Fig.\ref{fig2} shows the spin wave spectra with and without the non-linear correction. The difference is found to be small and is maximum at the zone-boundaries.

In Fig.\ref{fig3}, we show the calculated temperature dependences of magnon dispersion and sublattice magnetization. Within the regime where LSW approximation is valid, magnon energies do not alter much but the sublattice magnetization ($M_S$) reduces rather sharply with temperature. $M_S-T$ curve in the bottom panel of Fig.\ref{fig3} points to a temperature around 4.31 K where the AF order ceases to exist. It indicates the upper cut-off beyond which a spin wave theory around the AF reference state is not possible\cite{stratos,kar}. This mean field result falls between the experimental estimate of the Neel temperature $T_N=3.75~K$ and the temperature $T_{max}=6.95~K$ up to which short range AF correlation is observed in this compound\cite{kini}.

  The system size used for numerical computation is large enough to capture any short range AF correlation and thus the difference of the calculated value of $T_N$=4.31 K from the experimental finding ($T_N$=3.75 K\cite{kini,yusuf}) can be due to the usual over-estimation that occurs in any mean-field calculation.

In order to calculate the magnetic susceptibility, we need to consider the coupling to the external field $B$ and the term $-g\mu_BB\sum S_i^z$ is added to the Hamiltonian. The resulting zero-field molar susceptibility, obtained using Kubo's method of linear response function\cite{kubo} is given by
\begin{align}
\chi&=N_Ag^2S(S+1)\mu_B^2\int_0^\beta <{\bar S}_T^z(\tau){\bar S}_T^z(0)>d\tau\nonumber\\
=&2N_Ag^2S(S+1)\mu_B^2\beta\sum_kexp(\beta\omega_k)/(exp(\beta\omega_k)-1)^2
\label{eq9}
\end{align}
where ${\bar S}_T^z=\sum_iS_i^z.$ 
This gives the low temperature limit of the susceptibility while at high temperature, Curie-Weiss behavior is followed and the susceptibility is given by,
\begin{equation}
\chi(T)=\chi_0+\frac{C}{T+\theta}
\label{eq10}
\end{equation}
where $C=N_A\mu_{eff}^2/3k_B$ is the Curie constant and the other parameters being\cite{kini} $\mu_{eff}=1.68~\mu_B$, $\chi_0=-7$x$10^{-5}$ emu/mol, $N_A$ is Avogadro's number and $\theta=6.38~K$, the Curie-Weiss temperature (see Fig.\ref{fig4}).
We should mention here that the susceptibility we calculate is actually the parallel susceptibility $\chi_{||}$ where the external magnetic field is along $\hat{z}$, the easy axis direction. The susceptibility obtained experimentally from a powdered sample has contributions from perpendicular susceptibility $\chi_{\perp}$ as well ($\chi_{powder}=(\chi_{||}+2\chi_{\perp})/3$)\cite{sus1,sus2} and thus it takes non-zero value even at zero temperature\cite{kini}.

Spin wave calculation, being a mean field one, has little finite size effect as compared to more accurate quantum monte carlo (QMC) results as far as the low temperature behavior is concerned\cite{sus1}. Low temperature susceptibility behavior can thus be easily complimented by simple spin wave calculations without the expense of going to large size lattices. The inset in Fig.\ref{fig4} shows the spin-spin correlation results obtained from combined spin wave calculations at low temperature and QMC results at high temperatures using QMC Stochastic series expansion (SSE) directed loop-update method, as used in Ref.\cite{sudipta}.

\section{Compound $Zn_2Ti_{.25}V_{.75}O(PO_4)_2$}
Substitution of some of the $V$ ions of the compound $Zn_2VO(PO_4)_2$ with non-magnetic $Ti$ ions results in the new compound $Zn_2Ti_xV_{1-x}O(PO_4)_2$. This newly formed compound shows interesting structural modification  especially for $x=1/4$ substitution where one $V$ out of four $V$ ions in the square plaquette is substituted by $Ti$\cite{sudipta}.
 Ab-initio calculation\cite{sudipta} found the new structure to have a monoclinic A2 symmetry with lattice parameters a = 12.851 $\AA$, b = 9.180 $\AA$ and c = 12.852 $\AA$. This computer-designed structure gives a spin model with interesting magnetic properties\cite{sudipta}.
 Fig.\ref{fig1} (right panel) schematically shows the underlying spin lattice containing 3 parallel layers of $3\times3$ lattice planes of the substituted compound. 
\begin{figure}
\begin{center}
\epsfig{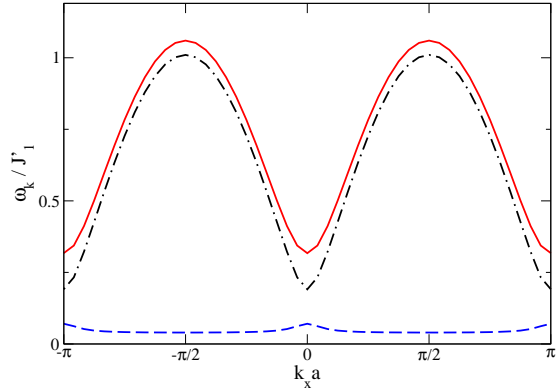}
\end{center}
\caption{Magnon modes of $Zn_2Ti_xV_{1-x}O(PO_4)_2$ for $k_y=k_z=0$. Six sublattices produce three sets of magnon modes (solid, dash-dotted and dashed lines). Here ``a'' denotes the lattice constant along $x$ direction for the substituted compound.}
\label{fig6}
\end{figure}

The lattice appears to be a repetition of two consecutive layers, only one of which has vacancies ($i.e.$, non-magnetic $Ti$ ion) in its alternate sites. The black and red (gray) balls in Fig.\ref{fig1} represent the crystal positions of magnetic $V$ ions and the vacancies are due to the presence of non-magnetic $Ti$ ions. Overall the system can be viewed as a sum of six different superlattices yielding three different sets of magnon modes each with degeneracy 2 ( because of the up-down symmetry). The strongest exchange interaction is found to be of AF type and along one bond direction within the $xy$ plane and it is much larger in value compared to the other intra-plane or inter-plane coupling\cite{sudipta}. The Hamiltonian for the new system in a $N_1\times N_2\times N_3$ cubic lattice (including both $V$ and $Ti$) is given by
\begin{eqnarray}
H&=&J_1^\prime\sum_{i=1}^{N_1}\sum_{j=1}^{N_2/2}\sum_{k=1}^{N_3} S_{i,2j,k}.S_{i+1,2j,k} + \nonumber\\&&J_1^{\prime\prime}\sum_{i=1}^{N_1/2}
\sum_{j=1}^{N_2/2}\sum_{k=1}^{N_3} S_{i^\prime,2j,k}.S_{i^\prime,2j\pm 1,k} - \nonumber\\&&J_c^\prime\sum_{i=1}^{N_1/2}\sum_{j=1}^{N_2/2}
\sum_{k=1}^{N_3} S_{i^\prime,2j,k}.S_{i^\prime,2j,k+1}-\nonumber\\
&&~~~~(i^\prime=2i-mod(k,2))
\label{eq11}
\end{eqnarray}

Within the LSW approximation and Fourier transformation, the Hamiltonian transforms to
\begin{align}
{H}&=E_{0}+2J_1^\prime S\sum_{k}[f_{11}(a_{1k}^\dagger a_{1k}+a_{4k}^\dagger a_{4k})+f_{22}(a_{2k}^\dagger a_{2k}\nonumber\\+&a_{3k}^\dagger a_{3k})+f_{55}(a_{5k}^\dagger a_{5k}+a_{6k}^\dagger a_{6k})+(cos(k_x)(a_{1k}a_{2k}\nonumber\\+&a_{3k}a_{4k})+\lambda^\prime (e^{ik_y}a_{1k}^\dagger a_{3k}+e^{-ik_y}a_{2k}^\dagger a_{4k})+h.c.)+\nonumber\\&\alpha^\prime\sum_{k}\{(a_{1k}^\dagger a_{1k}+a_{4k}^\dagger a_{4k})+(cos(k_z)(a_{1k}a_{6k}+\nonumber\\&a_{4k}a_{5k})+h.c.)\}]
\label{eq12}
\end{align}
where $a_{ik}$ represents the Fourier transformed bosonic spin-wave operators\cite{stratos} for the $i$-th superlattice, $f_{22}=1-\frac{\lambda^\prime}{2}$, $f_{11}=f_{22}+\alpha^\prime$, $f_{55}=\alpha^\prime$, $\alpha^\prime=J_1^{\prime\prime}/J_1^\prime$, $\lambda^\prime=J_c^{\prime}/J_1^\prime$ and $E_{0}$ is the vacuum or ground state energy. The wave-vector sums are within the 1st Brillouin zone corresponding to 3D unit cell (2a, 2b, 2c). With extended Bogoliubov transformation\cite{zhang}, applicable for this six sublattice system, the spin wave dispersions $\omega_k$ are obtained as a solution of the characteristic determinant equation : Det[$c_{ij}+\delta_{ij}(-1)^i\omega_{\textbf{k}}$]=0, where $c_{ij}$ represent coefficients or elements of the $6\times6$ Hamiltonian matrix in the basis of $a_{i,k}$'s.

Beyond the linear approximation, when quartic terms (within TDA) of the Hamiltonian are considered as well, the coefficients $c_{ij}$ gets modified to $\tilde{c}_{ij}$ with $\tilde{c}_{11}=\tilde{c}_{44}={c_{11}}-\frac{8}{N}\sum_q(\eta_q^{22}+\eta_q^{33}+\eta_q^{66})+(\xi_q^{12}cos(q_x)+\xi_q^{16}cos(q_y)+\xi_q^{13}cos(q_z)),\tilde{c}_{22}=\tilde{c}_{33}={c_{22}}-\frac{8}{N}\sum_q(\eta_q^{11}+\eta_q^{44}+\eta_q^{66})+(\xi_q^{12}cos(q_x)+\xi_q^{16}cos(q_y)+\xi_q^{13}cos(q_z)),\tilde{c}_{12}=\tilde{c}_{21}={c_{12}}-\frac{16}{N}\lambda\sum_q(\eta_q^{22}cos(k_x)+\xi_q^{12}cos(k_x-q_x)),\tilde{c}_{13}=\tilde{c}_{31}={c_{13}}-\frac{8}{N}\lambda\sum_q(\eta_q^{33}cos(k_z)+\xi_q^{13}cos(k_z-q_z)),\tilde{c}_{16}=\tilde{c}_{61}={c_{16}}-\frac{16}{N}\alpha\sum_q(\eta_q^{11}cos(k_y)+\xi_q^{16}cos(k_y-q_y)),\tilde{c}_{55}=\tilde{c}_{66}={c_{66}}-\frac{16}{N}\alpha\sum_q(\eta_q^{66}+\xi_q^{16}cos(q_y))$ the rest of the coefficients being zero. The correlation functions are obtained as
\begin{align}
\eta_{k}^{ii}&=\underset{\epsilon\rightarrow 0}{lim}~ i\int \frac{G_{ii}(k,w+i\epsilon)-G_{ii}(k,w-i\epsilon)}{e^{\beta w}-1}dw,\nonumber\\
\xi_{k}^{ij}&=\underset{\epsilon\rightarrow 0}{lim}~i\int\frac{G_{ij}(k,w+i\epsilon)-G_{ij}(k,w-i\epsilon)}{e^{\beta w}-1}dw
\label{eq13}
\end{align}
while Green's functions $G_{ij}(k,w)$'s and consecutively, coefficients $\tilde{c}_{ij}$ are obtained self-consistently by solving the matrix equations. Small parameter $\epsilon$ is taken to be $0.01dw$, where $dw=0.02J_1^\prime$ is the numerical resolution along the energy axis $w$. We find the spin wave modes by solving the modified characteristic equation. This could have been obtained identifying the singularities of $G_{ij}(k,w)$'s as well.
\begin{figure}[h]
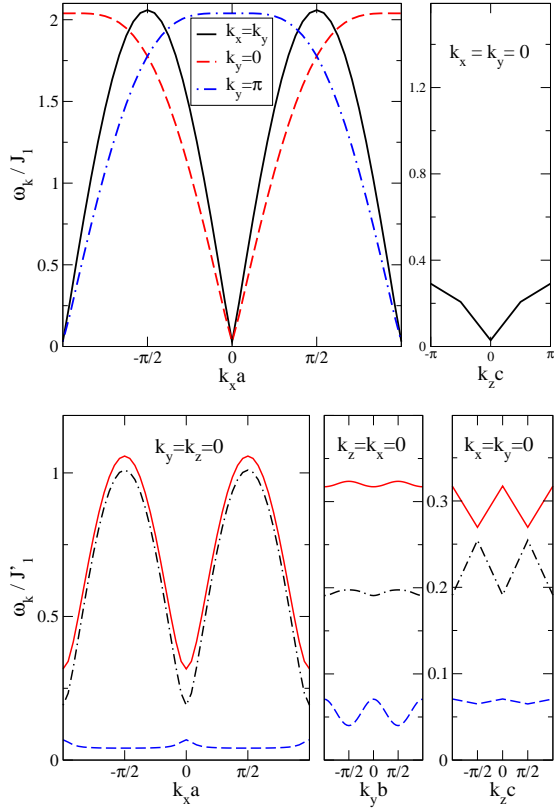

\begin{center}
\epsfig{file=old-model-allspwv3_new.eps,width=.9\linewidth,clip=}\\\vskip .1 in
\epsfig{file=new-model-allspwv2_new.eps,width=.9\linewidth,clip=}
\end{center}
\caption{Magnon modes of pristine (top) and substituted (bottom) compound. [ $k_z=0$ in the left figure of the top panel.] }
\label{fig7}
\end{figure}
\begin{figure}
\begin{center}
\epsfig{file=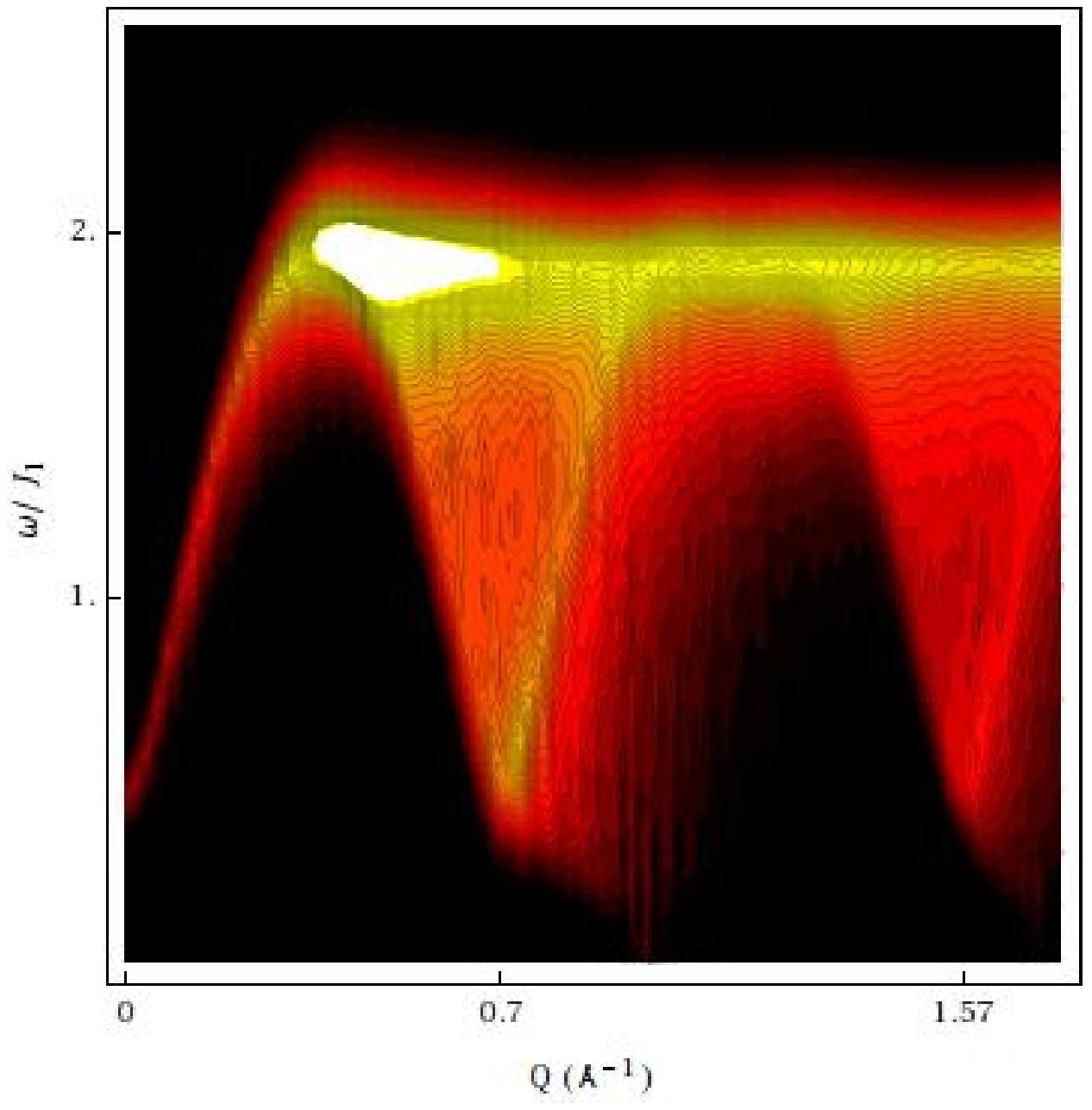,width=.9\linewidth,clip=}\\
\epsfig{file=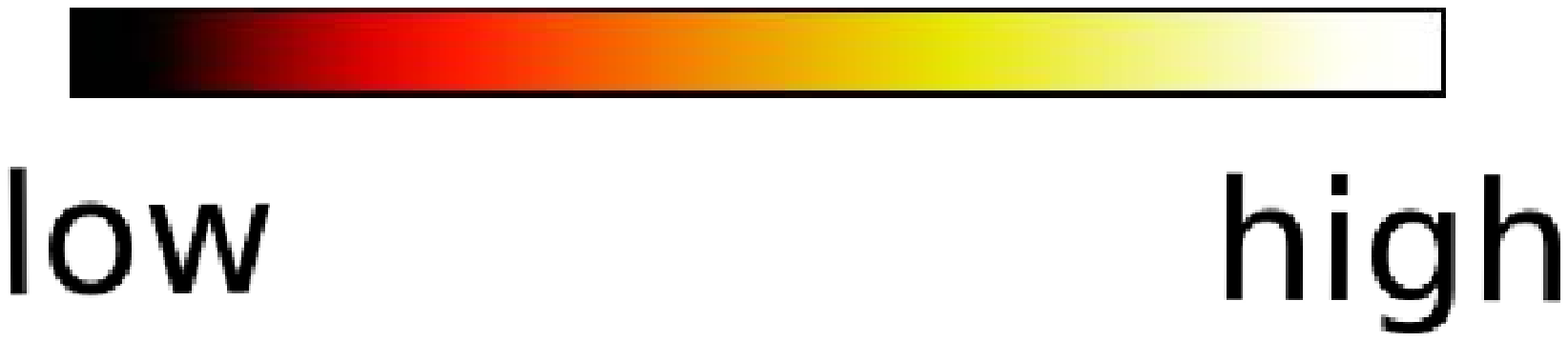,angle=-90,width=.6\linewidth,clip=}\\
\epsfig{file=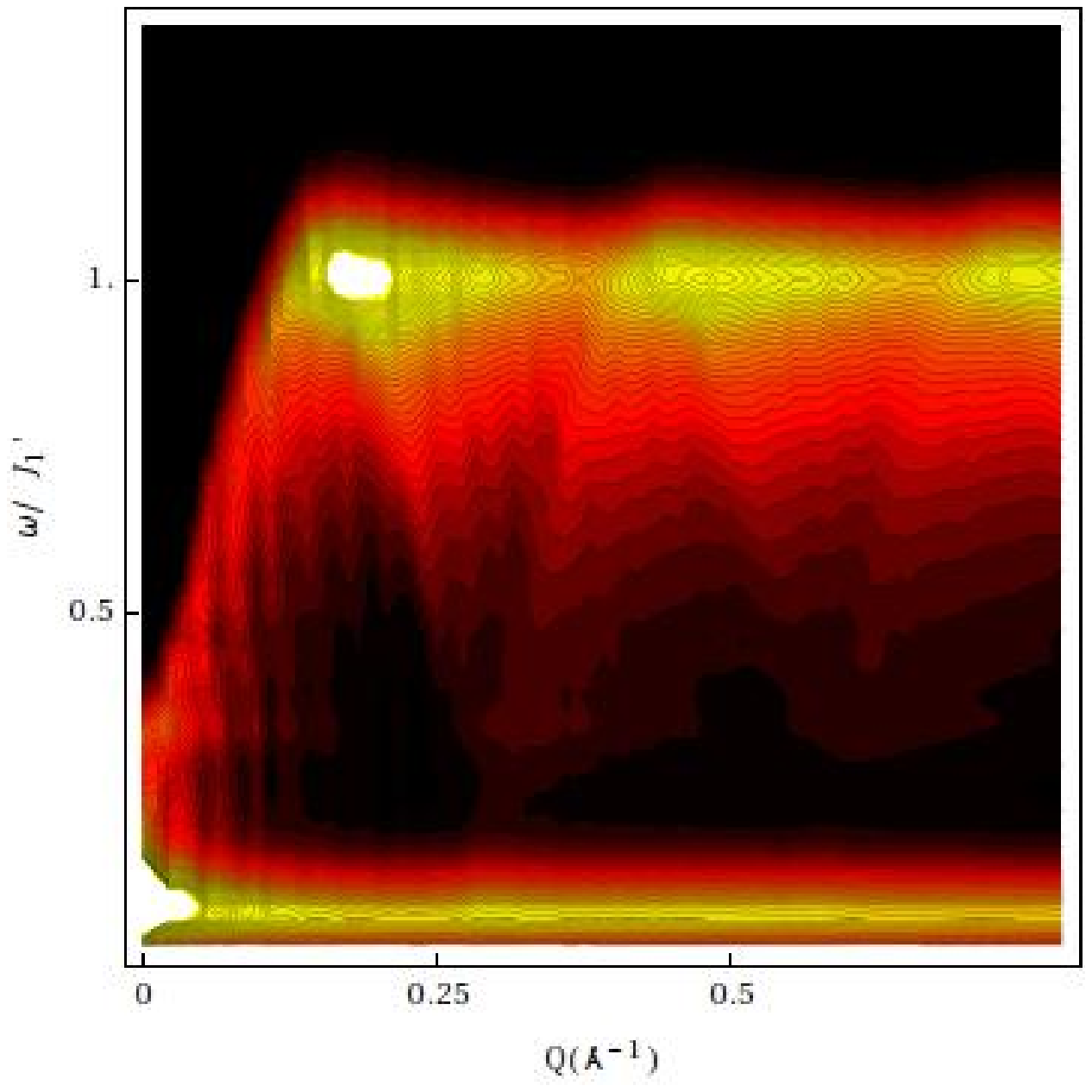,width=.9\linewidth,clip=}
\end{center}
\caption{(Color online) Powder averaged structure factor $S(Q,\omega)$ vs. $Q$ for the pristine (top) and substituted (bottom) compound calculated using eq.\ref{eq14} of Sec.5. Also see the discussion therein.}
\label{fig8}
\end{figure}
Fig.\ref{fig6} shows the spin wave dispersions along $k_x$ directions for $k_y=k_z=0$ in a $48\times48\times4$ lattice. Though we figure out the magnetic behavior of the compound to be of a quasi-1D AF, we keep the lattice size same as that of the pristine compound in order to compare the distinctive features between them. The different neighborhood of the three sublattices results in three different modes for this lattice. Weak coupling for the spins at sublattice E and F yields the magnon mode of small energy as compared to the energies of the other two modes.

\section{Comparison of Magnon Spectra}

Further look at the magnon modes for the pristine and substituted compound make the quasi-2D and quasi-1D feature in them very evident. Fig.\ref{fig7} shows the spin wave modes along different directions in k-space for both the compounds. The pristine compound shows large-scale variation in magnon energies along $x$ and $y$ (and not $z$) directions (Fig.\ref{fig7} top panels). But in the substituted compound, dispersion is visible mainly along $x$ direction (Fig.\ref{fig7}, bottom panels).

The inelastic neutron scattering cross-section of a system can be measured from the the magnetic structure factor $S(Q,\omega)$\cite{stratos,str-fac} which, within the approximation of one-magnon excitation (creation or annihilation), can be written as
\begin{align}
S({\bf{Q}},\omega)&\sim f(Q)^2\sum_\alpha |<0| S_q^{\perp}|\alpha>|^2\delta(\omega\pm E_\alpha)\nonumber\\&=f(Q)^2\frac{\omega_q^{max}\pm\sqrt{(\omega_q^{max})^2-\omega_q^2}}{\omega_q}\delta(\omega\pm\omega_q).
\label{eq14}
\end{align}
Here $S^{\perp}$ denotes $S^x$ and $S^y$, the spin components transverse to the easy $z$ direction. ${\bf{Q}}$ represents the momentum transfer of the inelastic scattering and is related to the BZ vector ${\bf{q}}$ as ${\bf{Q}}={\bf{G}}+{\bf{q}}$, ${\bf{G}}$ being a reciprocal lattice vector. $E_\alpha$ is the energy of the eigenstate $|\alpha>$ and $f({\bf{Q}})$ represents the magnetic form factor. $f({\bf{Q}})$ value is taken from an analytic expression (obtained using dipole approximation) with coefficients computed for the ion $V^{4+}$\cite{form factor}. Fig.\ref{fig8} shows the plot of the powder averaged structure factor $S(Q,\omega)$ vs. $Q$ for our pristine (top panel) and substituted (bottom panel) compounds. These structure factor plots can be compared with powder-neutron diffraction data. For the pristine compound ($i.e., Zn_2VO(PO_4)_2$), the intensity plot shows high values for scattering cross-section spanned over almost the whole energy range at $Q\sim0.7~\AA^{-1}$ and $Q\sim1.57~\AA^{-1}$. They indeed correspond to magnetic Bragg peak positions (100) and (120) respectively, as observed in Neutron diffraction measurements of powder $Zn_2VO(PO_4)_2$\cite{yusuf}.

\begin{figure}
\begin{center}
\begin{minipage}[t]{0.65\linewidth}
\raisebox{-4cm}{\epsfig{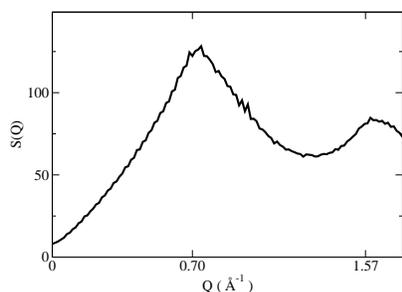}}
\end{minipage}\hfill
\begin{minipage}[t]{0.20\linewidth}
\caption{Integrated powder averaged structure factor S(Q) vs. Q for $Zn_2VO(PO_4)_2$.\label{fig9}}
\end{minipage}
\end{center}
\end{figure}

To make the comparison with experiment more quantitative, in Fig.\ref{fig9} we show the integrated powder averaged structure factor $S(Q)$ for the spin model of $Zn_2VO(PO_4)_2$, which can be obtained from integrating $S(Q,\omega)$ (Fig\ref{fig8}) over $\omega$. It indicates peaks in the scattering cross-section at $Q$ = 0.7 and 1.57 $\AA^{-1}$ corresponding to the magnetic Bragg peak positions (100) and (120) respectively. Same positions have been found by Rietveld refined powder neutron diffraction (PND) results on $Zn_2VO(PO_4)_2$, as shown in Fig.1c of Ref.\cite{yusuf}. There the magnetic pattern at a low temperature of 1.5 K (which is below the experimental $T_N=3.75$ K) is obtained by subtracting the corresponding PND pattern from the PND pattern at a high temperature of 300 K (which has only nuclear and no magnetic contributions).

In the compound $Zn_2Ti_{.25}V_{.75}O(PO_4)_2$, three magnon modes give three terms in the expression for $S(Q,\omega)$. The low energy mode is almost dispersion-less compared to the other two almost degenerate modes and thus gives featureless large scattering cross-section at low energies (Fig.\ref{fig8}, bottom panel). On the other hand, the higher energy nearly degenerate two other magnon modes are responsible for the diffused magnetic scattering peaks at values of Q which are smaller compared to that for the pristine compound. This is because of the larger direct-space unit cell in the substituted compound as compared to the pristine one. The overall $S(Q,\omega)$ distribution of the substituted compound, unlike the pristine one, lacks sharpness in peaks indicating strengthening of the quantum fluctuation effect due to the reduced dimensionality of the spin Hamiltonian.

\section{Summary and Discussion}
To summarize, we have presented a linear and non-linear spin wave study for a quasi-2D AF with parameters representing the compound $Zn_2VO(PO_4)_2$. The damping of the spin wave modes with temperature is observed and a decay in sublattice magnetization leading to a $T_{max}=4.31~K$ is witnessed. This temperature is high compared to the experimentally obtained $T_N=3.75~K$. This difference can be a result of the approximation used in our calculation and can be improved using 1/N expansion of the nonlinear-$\sigma$ model as shown for quasi-2D antiferromagnet by Irkhin et al.\cite{Irkhin}. The parallel susceptibility at zero field is obtained at low temperature which can well complement QMC data that shows enough finite size effect at low T. 

Further a spin wave study for the $Ti$-substituted compound $Zn_2Ti_{.25}V_{.75}O(PO_4)_2$ shows the multiple magnon modes obtained as a result of structural modification. The six superlattice structure creates small size BZ in the system which is also evident from the structure factor plot (Fig.\ref{fig8}). The magnon dispersions also confirm the quasi-1D nature of the underlying spin lattice.

We should mention here that our computer-designed compound $Zn_2Ti_{.25}V_{.75}O(PO_4)_2$ has an ordered sublattice structure between V and Ti. Our model, therefore, represents an idealistic situation where any possible interchange of the Ti and V ion occupied sites, that may occur in actual synthesis of such compound, is not considered. 
The possibility of such inter-mixing, however, cannot be ruled out. In future, we would like to address this aspect in our theoretical study.

\section{Acknowledgment}
S. Kar thanks Dr. Anup Bera for fruitful discussion on Neutron powder diffraction calculation. T.S.D. would like to acknowledge support from DST.

\end{document}